\documentclass[11pt]{article}

\usepackage[utf8]{inputenc}
\usepackage[T1]{fontenc}
\usepackage[margin=1in]{geometry}
\usepackage{abstract}
\usepackage{titlesec}
\usepackage{enumitem}
\usepackage{hyperref}
\usepackage{xcolor}

\hypersetup{
  colorlinks=true,
  linkcolor=black,
  citecolor=black,
  urlcolor=blue,
  pdftitle={Weather- and Location-Aware Agentic Dining Recommendation},
  pdfauthor={Kadharmoideen Fadurudeen}
}

\titlespacing*{\section}{0pt}{1.4ex plus 1ex minus .2ex}{1ex plus .2ex}
\titlespacing*{\subsection}{0pt}{1.1ex plus 1ex minus .2ex}{0.8ex plus .2ex}

\title{\textbf{Weather- and Location-Aware Agentic Dining Recommendation: Leveraging LLM World Knowledge for Region-Sensitive Contextual Reasoning}}

\author{
  Kadharmoideen Fadurudeen\\
  Independent Researcher\\
  \texttt{kadhar.mfn@gmail.com} \quad \url{https://kadhar.dev}
}
\date{}

\begin{document}
\maketitle

\begin{abstract}
\noindent
Context-aware recommender systems have long recognized that factors such as location, time, and weather shape where and what people choose to eat. Existing weather-aware food and point-of-interest recommenders, however, typically treat weather generically---mapping conditions to preferences through hand-crafted rules or specially trained context models---and do not capture that the \emph{culturally appropriate} response to weather is itself region-specific: a rainy evening calls for hot tea and fried snacks in one culinary culture and for very different comfort food in another. Encoding such weather-by-region-by-cuisine interactions as explicit rules or training data is brittle and does not scale. We present a weather- and location-aware agentic dining-recommendation system that takes a different approach: a large language model (LLM) orchestrates tools for location and weather retrieval and then \emph{reasons in natural language} over the combined context, drawing on the cultural and culinary world knowledge already latent in the model to produce region-sensitive, weather-appropriate recommendations without per-region rule tables or specialized training. We describe the agent architecture, the tool-orchestration flow (Google location services and a weather service feeding an OpenAI LLM), and the reasoning mechanism, and we report on a working prototype that was implemented and briefly deployed end-to-end. We discuss design trade-offs---cost, latency, ambiguity handling, and fallbacks---and we are explicit about limitations, including the absence of a formal user study and the risk of cultural stereotyping in locality-based inference. The contribution is architectural: a simple, extensible pattern for incorporating environmental and cultural context into agentic recommendation through LLM reasoning rather than engineered rules.

\vspace{0.6em}
\end{abstract}

\section{Introduction}

Choosing where and what to eat is a context-dependent decision. The same person wants different things on a hot afternoon than on a cold, rainy evening; near the office at lunchtime than at home late at night. Recommender-systems research has recognized this for well over a decade, and context-aware recommender systems (CARS) routinely incorporate signals such as location, time of day, companionship, and weather to improve suggestions.

Yet most weather-aware food and point-of-interest recommenders treat weather as a \emph{generic} signal: rain favors indoor venues, heat favors cold items, and so on. This overlooks a subtlety that matters in practice: the culturally appropriate response to a given weather condition is itself region-specific. On a rainy evening in parts of South India, a common craving is hot tea with fried snacks such as vada or bajji; in much of the United States, rainy-day comfort food might instead be pizza or soup. On a hot day, relief in South India often means buttermilk, lime juice, or tender coconut water, whereas elsewhere it might mean iced soft drinks or ice cream. The mapping is not weather-to-food; it is \emph{weather-and-place}-to-food.

Capturing these interactions with the traditional toolkit is hard. Rule-based systems would require analysts to hand-author weather-by-region-by-cuisine mappings for every locale; trained context models would require labeled data spanning the same combinatorial space. Both are brittle and expensive to extend to new regions.

We observe that large language models (LLMs) already encode broad cultural and culinary world knowledge, and that this knowledge can be exploited directly. Instead of engineering weather-to-food rules, we let an LLM agent \emph{reason} over the combined context---the user's location (from which regional culinary norms can be inferred) and the current weather---and produce recommendations accordingly. This turns a data-and-rules engineering problem into a reasoning problem that the model is already equipped to handle.

This paper makes the following contributions:

\begin{enumerate}[leftmargin=1.4em]
  \item An \textbf{agentic architecture} for dining recommendation in which an LLM orchestrates location and weather tools and reasons over their combined output.
  \item A \textbf{weather- and location-aware reasoning mechanism} that leverages the LLM's latent cultural knowledge to produce region-sensitive, weather-appropriate recommendations \emph{without} hand-crafted rules or specialized training.
  \item An honest account of a \textbf{working prototype} that was implemented and briefly deployed end-to-end, together with a discussion of design trade-offs and limitations---including cost, latency, and the risk of cultural stereotyping in locality-based inference.
\end{enumerate}

Our claim is deliberately narrow and, we believe, defensible: weather-aware recommendation is not new, but using an LLM agent's world knowledge to perform \emph{region-sensitive} weather-aware reasoning, in place of engineered context models, is an under-explored and practically attractive alternative.

\section{Related Work}

\paragraph{Context-aware recommendation.}
The incorporation of contextual information---beyond user and item attributes---into recommendation is a well-established field. Surveys of context-aware recommender systems describe location, time, season, temperature, companionship, and weather as standard contextual dimensions, and show that incorporating such context can significantly improve recommendation quality in domains including restaurants~\cite{adomavicius2011}. Our work sits squarely within this tradition; we do not claim context-awareness, or weather-awareness, as novel.

\paragraph{Weather-aware food and POI recommendation.}
Weather has specifically been studied as a contextual signal. Production systems have exploited weather among other contextual features: a drive-thru recommendation model deployed at a major fast-food chain uses location, time, and weather via learned representations to recommend menu items~\cite{gao2020}. Research on the impact of weather for point-of-interest recommendation has examined how conditions affect where people choose to go~\cite{trattner2016}. These approaches typically rely on feature engineering or trained models. Our contribution differs in \emph{mechanism}: we use an LLM's natural-language reasoning over weather rather than learned or hand-crafted weather features, and we focus on the region-specific interaction between weather and cuisine, which generic weather features do not capture.

\paragraph{LLMs, tool use, and agentic workflows.}
Recent systems use LLMs as agents that plan and call external tools to accomplish tasks, deciding at run time which tool to invoke rather than following a fixed script. Work integrating LLMs with live data sources---for example combining a mapping/places API with a GPT-class model to provide recommendations grounded in real-time business data---demonstrates both the promise and the engineering considerations of this pattern~\cite{smart2025}. We adopt this agentic, tool-using paradigm and apply it to weather- and location-aware dining recommendation.

\paragraph{Positioning.}
Our contribution is therefore not that weather matters, nor that context-aware recommendation is possible, but that an LLM agent can subsume the role of engineered weather/culture models by reasoning over contextual signals directly, using knowledge it already contains. To our knowledge this specific framing---LLM-reasoned, region-sensitive weather-aware dining recommendation within a tool-orchestration loop---has received little direct attention.

\section{System Architecture}

The system is a tool-using LLM agent. A natural-language query (for example, ``find me a good lunch place nearby'') triggers an orchestration loop in which the LLM decides which tools to call, invokes them, and reasons over their results to compose a recommendation.

The agent has access to three categories of tool:

\begin{itemize}[leftmargin=1.4em]
  \item \textbf{Location services.} Google location/places APIs resolve the user's location (coordinates and locality) and retrieve candidate nearby restaurants with attributes such as name, address, rating, price level, and opening status.
  \item \textbf{Weather retrieval.} A weather service (OpenWeather) provides current conditions---temperature, precipitation, and general weather state---for the resolved location.
  \item \textbf{Restaurant/candidate search.} Structured search over candidate venues by criteria such as cuisine, meal type, price range, features (e.g., buffet), and radius.
\end{itemize}

The orchestration follows the tool-calling pattern: the LLM is given the user query and the available tool schemas; it calls the location tool, then the weather tool, then the restaurant-search tool, accumulating context across steps; and it finally reasons over the aggregated context---location/region, current weather, and candidate restaurants---to produce a ranked, explained recommendation. Multi-step reasoning is bounded by a step limit to control cost and latency.

\section{Weather- and Location-Aware Reasoning}

The novel component of the system is how the LLM uses the combined location and weather context.

\paragraph{Location as a regional/cultural signal.}
The location returned by the mapping API is not merely a coordinate for proximity search; it also serves as a proxy for the regional culinary context. The locality is passed to the LLM, which can infer likely regional cuisine norms associated with that place. This inference is performed entirely by the model's latent knowledge---no explicit region-to-cuisine table is maintained by the system.

\paragraph{Weather as a comfort/appropriateness signal.}
The current weather---temperature band, precipitation, and time of day---is passed to the LLM as structured natural-language context. Rather than applying fixed rules (e.g., ``if temperature $>$ 85\,\textdegree F, prefer cold items''), the system lets the model reason about what is appropriate.

\paragraph{Joint reasoning.}
Crucially, the model reasons over location and weather \emph{together}, which is where region-sensitive behavior emerges. For example, given a rainy evening, the model may lean toward hot tea and fried snacks in a South Indian locality but toward different comfort foods elsewhere; given a hot afternoon, it may suggest buttermilk or tender-coconut options in one region and iced beverages in another. These behaviors are not encoded anywhere in the system; they are elicited from the model's world knowledge by supplying the right context. This is the core argument of the paper: the combinatorial weather-by-region-by-cuisine space that would be impractical to hand-engineer is already, approximately, represented in a capable LLM and can be accessed through reasoning.

We stress that these are illustrative of the system's observed behavior, not validated claims about optimal recommendations; Section~\ref{sec:limits} discusses the limitations of relying on locality-based cultural inference.

\section{Implementation and Prototype}

The prototype was implemented as a full-stack application: an LLM agent (OpenAI GPT-4-class model) using a tool-calling framework to orchestrate the Google location/places services, the OpenWeather service, and restaurant candidate search, with a lightweight front end for entering queries and viewing recommendations.

The system was built and run \textbf{end-to-end}---from natural-language query, through location resolution and weather retrieval, to LLM reasoning and a returned, explained recommendation---and was briefly \textbf{deployed live}. It was subsequently taken offline to avoid the ongoing costs of the paid location APIs. The prototype therefore constitutes a demonstrated implementation of the architecture rather than a currently-running service. Its purpose is to establish feasibility of the approach and to surface the design considerations discussed next, not to serve as an evaluated production system.

\section{Design Considerations}

Building the prototype surfaced several practical considerations that generalize to agentic recommendation systems:

\begin{itemize}[leftmargin=1.4em]
  \item \textbf{API cost control.} External APIs---particularly location/places services and per-token LLM calls---dominate operating cost. Caching location and weather results for repeated or nearby queries, and bounding the number of agent reasoning steps, materially reduce cost. Indeed, ongoing API cost was the reason the live deployment was retired.
  \item \textbf{Latency.} Each tool call and LLM step adds latency. Limiting steps, parallelizing independent tool calls where possible, and caching keep response times acceptable.
  \item \textbf{Ambiguity handling.} When a query is under-specified, the agent can either ask a clarifying question or proceed with sensible defaults; the right choice depends on the interaction context.
  \item \textbf{Fallbacks.} When no candidate matches the inferred criteria, the agent should broaden the search or suggest nearby alternatives rather than returning nothing.
\end{itemize}

\section{Limitations and Future Work}
\label{sec:limits}

The prototype demonstrates feasibility but has not been formally evaluated; we report no user study or quantitative comparison against baseline recommenders. Rigorous evaluation---measuring recommendation quality, user satisfaction, and the value added by region-sensitive reasoning relative to generic weather-awareness---is the primary avenue for future work.

Several specific limitations warrant emphasis. First, \textbf{locality is not preference}: inferring regional cuisine norms from a user's location is a heuristic that can be wrong for any individual---a person in one region may prefer another region's cuisine---so location-based cultural inference should inform, not dictate, recommendations. Second, \textbf{LLM cultural knowledge is uneven and can stereotype}: the model's associations between place, weather, and food may be inaccurate, coarse, or biased, particularly for less-represented regions, and region-sensitive reasoning therefore requires validation and guardrails. Third, the system depends on \textbf{external paid APIs}, with the attendant cost and availability constraints noted above.

Future work includes personalization (combining inferred regional context with individual user history), forecast-aware planning (reasoning over upcoming rather than only current conditions), systematic evaluation of the cultural-reasoning behavior across diverse regions, and mitigation of cultural bias in the model's outputs.

\section{Conclusion}

Weather has long been recognized as a useful context for food and venue recommendation, but the culturally appropriate response to weather is region-specific, and encoding that interaction with hand-crafted rules or trained models is brittle and hard to scale. We presented a weather- and location-aware agentic dining recommender that instead lets an LLM reason over combined location and weather context, drawing on the cultural and culinary knowledge already latent in the model to produce region-sensitive, weather-appropriate recommendations. A working prototype, implemented and briefly deployed end-to-end, demonstrates the feasibility of the approach, and our discussion of cost, latency, and---importantly---the risks of locality-based cultural inference is intended to help others build on the idea responsibly. The contribution is architectural: a simple, extensible pattern for bringing environmental and cultural context into agentic recommendation through reasoning rather than engineered rules.

\section*{Acknowledgments}
This work was developed as an independent personal project.


\begin{thebibliography}{9}
\bibitem{adomavicius2011}
G.~Adomavicius and A.~Tuzhilin.
\newblock Context-Aware Recommender Systems.
\newblock In \emph{Recommender Systems Handbook}, pages 217--253. Springer, 2011.
\newblock DOI: 10.1007/978-0-387-85820-3\_7.

\bibitem{trattner2016}
C.~Trattner, A.~Oberegger, L.~Eberhard, D.~Parra, and L.~B.~Marinho.
\newblock Understanding the Impact of Weather for POI Recommendations.
\newblock In \emph{Proc. Workshop on Recommenders in Tourism (RecTour) at ACM RecSys}, Boston, MA, USA, pages 16--23. CEUR-WS Vol.\ 1685, 2016.

\bibitem{gao2020}
L.~Wang, K.~Huang, J.~Wang, S.~Huang, J.~Dai, and Y.~Zhuang.
\newblock Context-Aware Drive-thru Recommendation Service at Fast Food Restaurants.
\newblock \emph{arXiv preprint} arXiv:2010.06197, 2020.
\newblock (Transformer Cross Transformer (TxT); deployed at Burger King, using location, time, and weather.)

\bibitem{smart2025}
SMART Restaurant ReCommender: A Context-Aware Restaurant Recommendation Engine.
\newblock \emph{AI (MDPI)}, 6(4):64, 2025.
\newblock DOI: 10.3390/ai6040064.
\end{thebibliography}
\end{document}